\documentclass[aps,pre,twocolumn,superscriptaddress,amsmath,amssymb,final,letterpaper,10pt,longbibliography]{revtex4-1}

\usepackage{graphicx}
\usepackage[hang]{subfigure}
\usepackage{epstopdf}
\usepackage{times}
\usepackage{bm}
\usepackage{xcolor}
\usepackage{subfigure}
\usepackage{microtype}
\usepackage{hyperref}
\hypersetup{colorlinks=true,linkcolor=black,citecolor=black,urlcolor=black}
\hypersetup{pdfauthor={Jean-Gabriel Young}}
\hypersetup{pdfsubject={Structural preferential attachment revisited}}

\begin{document}
\title{Growing networks of overlapping communities with internal structure}
\author{Jean-Gabriel Young}
\affiliation{D\'epartement de Physique, de G\'enie Physique, et d'Optique, Universit\'e Laval, Qu\'ebec (Qu{\'e}bec), Canada, G1V 0A6}
\author{Laurent H\'ebert-Dufresne}
\affiliation{D\'epartement de Physique, de G\'enie Physique, et d'Optique, Universit\'e Laval, Qu\'ebec (Qu{\'e}bec), Canada, G1V 0A6}
\affiliation{Santa Fe Institute, Santa Fe, New Mexico 87501, USA}
\author{Antoine Allard}
\affiliation{D\'epartement de Physique, de G\'enie Physique, et d'Optique, Universit\'e Laval, Qu\'ebec (Qu{\'e}bec), Canada, G1V 0A6}
\affiliation{Departament de F\'isica de la Mat\`eria Condensada, Universitat de Barcelona, Mart\'i i Franqu\`es 1, E-08028 Barcelona, Spain}
\affiliation{Universitat de Barcelona Institute of Complex Systems (UBICS), Universitat de Barcelona, Barcelona, Spain}
\author{Louis J. Dub\'e}
\affiliation{D\'epartement de Physique, de G\'enie Physique, et d'Optique, Universit\'e Laval, Qu\'ebec (Qu{\'e}bec), Canada, G1V 0A6}
\begin{abstract}
We introduce an intuitive model that describes both the emergence of community structure and the evolution of the internal structure of communities in growing social networks.
The model comprises two complementary mechanisms: One mechanism accounts for the evolution of the internal link structure of a single community, and the second mechanism coordinates the growth of multiple overlapping communities.
The first mechanism is based on the assumption that each node establishes links with its neighbors and introduces new nodes to the community at different rates.
We demonstrate that this simple mechanism gives rise to an effective maximal degree within communities.
This observation is related to the anthropological theory known as Dunbar's number, i.e., the empirical observation of a maximal number of ties which an average individual can sustain within its social groups.
The second mechanism is based on a recently proposed generalization of preferential attachment to community structure, appropriately called structural preferential attachment (SPA).
The combination of these two mechanisms into a single model (SPA+) allows us to reproduce a number of the global statistics of real networks: The distribution of community sizes, of node memberships and of degrees.
The SPA+ model also predicts (a) three qualitative regimes for the degree distribution within overlapping communities and (b) strong correlations between the number of communities to which a node belongs and its number of connections within each community.
We present empirical evidence that support our findings in real complex networks.
\end{abstract}
\maketitle

%==============================================================================
\section{Introduction}
\label{sec:introduction}
%==============================================================================
Networks are at the center of the quantitative analysis of social systems \cite{Wasserman1994}.
They encode the social ties among different individuals within a mathematical construct that allows a quantitative assessment of the role of individuals in social networks through various measures, and the analysis of correlations among them \cite{Wasserman1994,Newman2010}. 
One instance of these correlations, the similarity between the neighborhoods of different nodes (individuals), has received particular attention since links tend to be clustered in tightly connected groups \cite{Watts1998,Girvan2002}.
Networks are often expressed as a superposition of such densely connected groups, and we refer to this decomposition as the community structure of a network \cite{Fortunato2010, Xie2013}.

We consider the problem of modeling both the emergence of community structure in social networks and the growth of the internal structure of these communities. 
Many community detection algorithms and community modeling efforts consider a fully random, or Erd\H{o}s-R\'{e}nyi (ER), internal structure \cite{Peixoto2012,Clauset2008,Guimera2009,Seshadhri2012,Yang2012}.
This is a principled approach, in the sense that it relies on minimal \textit{a priori} information, but it is unfortunately incompatible with most common growth processes in two respects.
One, it ignores the temporal aspect of community growth \cite{HebertDufresne2016}.
Two, it ignores the fact that nodes can have very heterogeneous structural roles in complex networks \cite{Barabasi1999}.

The preferential attachment mechanism (PA) \cite{simon1955class, price1976general, Barabasi1999} offers a simple way to include the temporal and heterogeneous aspects of complex networks in growth processes.
PA is based on the assumption that a node's current state is a good indicator of its future behavior.
We take inspiration from the PA model \cite{Barabasi1999} and its recent extension to community structure \cite{HebertDufresne2011,HebertDufresne2012}.
We combine heterogeneous PA at the level of communities with minimal \textit{a priori} information for the internal structure of communities.
That is, we postulate simple rules for the growth of the internal structure of communities.
In so doing, we provide a new growth process that reproduces a number of important properties of overlapping community structures and complex networks.

The structure of the paper is as follows.
In Sec.~\ref{sec:internal_model}, we describe a  process by which a single community and its structure may grow. We find an upper bound on how many connections an average individual may maintain as the community grows. This finding is discussed in relation to the anthropological theory known as Dunbar's number.
In Sec.~\ref{sec:complete_model}, we incorporate this internal community growth process within a preferential attachment model at the community structure level and provide a recipe for its implementation. 
This yields a general model for the concurrent growth of overlapping and heterogeneous communities.
In Sec.~\ref{sec:results}, we compare our model to empirical data and investigate its implications.
We find that our model generates networks whose global statistics are comparable to that of real networks, and that their internal community structure contain correlations also present in empirical datasets.
We close with a short conclusion in Sec.~\ref{sec:conclusion}, and relegate some of the technical details to two Appendices.

%==============================================================================
\section{Growth of a single community}
\label{sec:internal_model}
%==============================================================================
In this first section, we introduce a simple model that describes the growth of a \emph{single} community, independently of the rest of the network.
The model builds on the recent observation that the rate of growth of a community is predicted by preferential attachment \cite{HebertDufresne2011,HebertDufresne2012,HebertDufresne2015}.
This hypothesis is known to reproduce some of the statistical properties of the community structure of real networks \cite{HebertDufresne2011,HebertDufresne2012}.
It can be interpreted as if each node in a community introduces new nodes at a fixed rate: The more nodes, the faster the community grows with respect to other competing communities in the same network.
In what follows, we combine this node creation mechanism to an elementary link creation mechanism, and obtain a reasonable model for the growth of a single community.

%~~~~~~~~~~~~~~~~~~~~~~~~~~~~~~~~~~~~~~~~~~~~~~~~~~~~~~~~~~~~~~~~~~~~~~~~~~~~~~
\subsection{Description of the model and mean-field analysis}
\label{subsec:internal_model-description_mean_field}
%~~~~~~~~~~~~~~~~~~~~~~~~~~~~~~~~~~~~~~~~~~~~~~~~~~~~~~~~~~~~~~~~~~~~~~~~~~~~~~
We model the growth of a single community with a continuous-time Markov process.
The model is simply stated.
A community is initially represented by a small graph, e.g. a triad or a single node.
Each of these nodes recruit new nodes at a constant rate $\rho_r$;
at time $t$, the growth rate $\dot{n}(t)$ of the community is therefore proportional to $\rho_rn(t)$, where $n(t)$ is the size of the community.
Whenever a new node is recruited, it is at first only connected to the node who recruited it (its degree $k$, i.e. number of neighbors, therefore equals $1$ within the community).
To allow for denser communities, we introduce another mechanism whereby each node initiates the creation of an undirected link at a constant rate $\rho_\ell$ (unless it is already connected to every node).
A second node is randomly selected to complete the link (note that we exclude self-loops and multiple links).

The average number  $m_k(t)$ of nodes with degree $1<k<n-1$ within an average community of size $n(t)$ can be followed through continuous time $t$ with the interdependent set of rate equations
\begin{subequations}
\label{eq_group:master_equations_internal}
\begin{align}
    \label{eq:compartemental_master_equation}
    \dot{m}_k(t) &= \rho_r\bigl(m_{k-1} - m_k\bigr)+\rho_\ell\bigl(m_{k-1} - m_k\bigr)\notag\\
                 &+ \rho_\ell X \Bigg[\frac{(\lfloor\!n\!\rfloor-k)m_{k-1}}{Z}- \frac{(\lfloor\!n\!\rfloor- k - 1)m_{k}}{Z}\Bigg] \; ,
\end{align}
where $\lfloor\!n\!\rfloor$ is the integer part of $n$, where $X := \sum_{k'=1}^{\lfloor\!n\!\rfloor-2}m_{k'}(t)$ is the number of nodes that can initiate link creation events, and where $Z := \sum_{k'=1}^{\lfloor\!n\!\rfloor-2}\left(\lfloor\!n\!\rfloor-1-k'\right)m_{k'}(t)$ is the total number of potential links.
The first term accounts for the arrival of new nodes: Each node recruits at rate $\rho_r$ and gains new connections accordingly.
This creates a flow that brings a node of degree $k-1$ to degree $k$ [positive effect on $m_k(t)$] and node of degree $k$  to degree $k+1$ [negative effect].
The second term is due to the creation of new links: Each node initiates the creation of a new link at rate $\rho_\ell$, and the net effect on $m_k(t)$ is identical to that of the node creation mechanism.
The third term accounts for the increase in degree incurred by a node randomly selected to complete new links.
Events of this type occur at rate $\rho_{\ell}X$ and affect nodes of degree $k$ with probability $(\lfloor\!n\!\rfloor-k-1)m_k/Z$.

Equation \eqref{eq:compartemental_master_equation} is only valid when $1<k<n-1$ for two reasons.
One, nodes of degrees $k=\lfloor\!n\!\rfloor-1$ cannot initiate or receive new links.
Two, node creation only involves nodes of degree $k=1$.
Another set of rate equations is therefore needed to handle the limit cases.
We find
\begin{align}
    \dot{m}_0(t) &= 0\\
    \dot{m}_1(t) &= \!\rho_r n \! - \rho_r m_1 \!-\! \rho_\ell\left[m_1 + \! X \frac{(\lfloor\!n\!\rfloor-2)m_1}{Z}\right] \label{eq:compartemental_master_equation_n1}\\
    \dot{m}_{\lfloor\!n\!\rfloor-1}(t) &= m_{\lfloor\!n\!\rfloor-2}\left(\rho_r  + \rho_\ell + \rho_\ell\frac{X}{Z}\right) - \rho_{r}m_{\lfloor\!n\!\rfloor-1}\label{eq:compartemental_master_equation_n_n_m1}\\
    \dot{m}_{\lfloor\!n\!\rfloor}(t) &= \rho_{r}n_{\lfloor\!n\!\rfloor-1} \; . \label{eq:compartemental_master_equation_n_n}
\end{align}
\end{subequations}
Note that the set of Eqs.~\eqref{eq_group:master_equations_internal} becomes inconsistent  when $\lfloor\!n\!\rfloor \leq 2$, since we obtain different equations for a same compartment $m_k(t)$.
Fortunately, we do not need Eqs.~\eqref{eq_group:master_equations_internal} to track the evolution of the community when $n\leq 2$ ---this evolution is deterministic.
A community that contains a single node must first grow: The only node is already of maximal degree, and link creation events never occur.
The same reasoning applies to the case of $n=2$.
Therefore, whenever $n(t_0) \leq 2$, we can instead use the initial condition $\vec{m}(t_0)=(m_0=0, m_1=2, m_2=1)$, i.e. track the community starting from the point where randomness plays a role.
If temporal information is important, then one can compute the expected amount of time spent in configurations of sizes $n\leq 2$, and correct the prediction {\em a posteriori} (the delay is an exponentially distributed random variable).

% - - - - - - - - - - - - - - - - - - - - - - - - - - - - - - - - - - - - - - -
% Internal degree distribution at different sizes and values of the ratio r
\begin{figure*}
    \centering
    \includegraphics[scale=1]{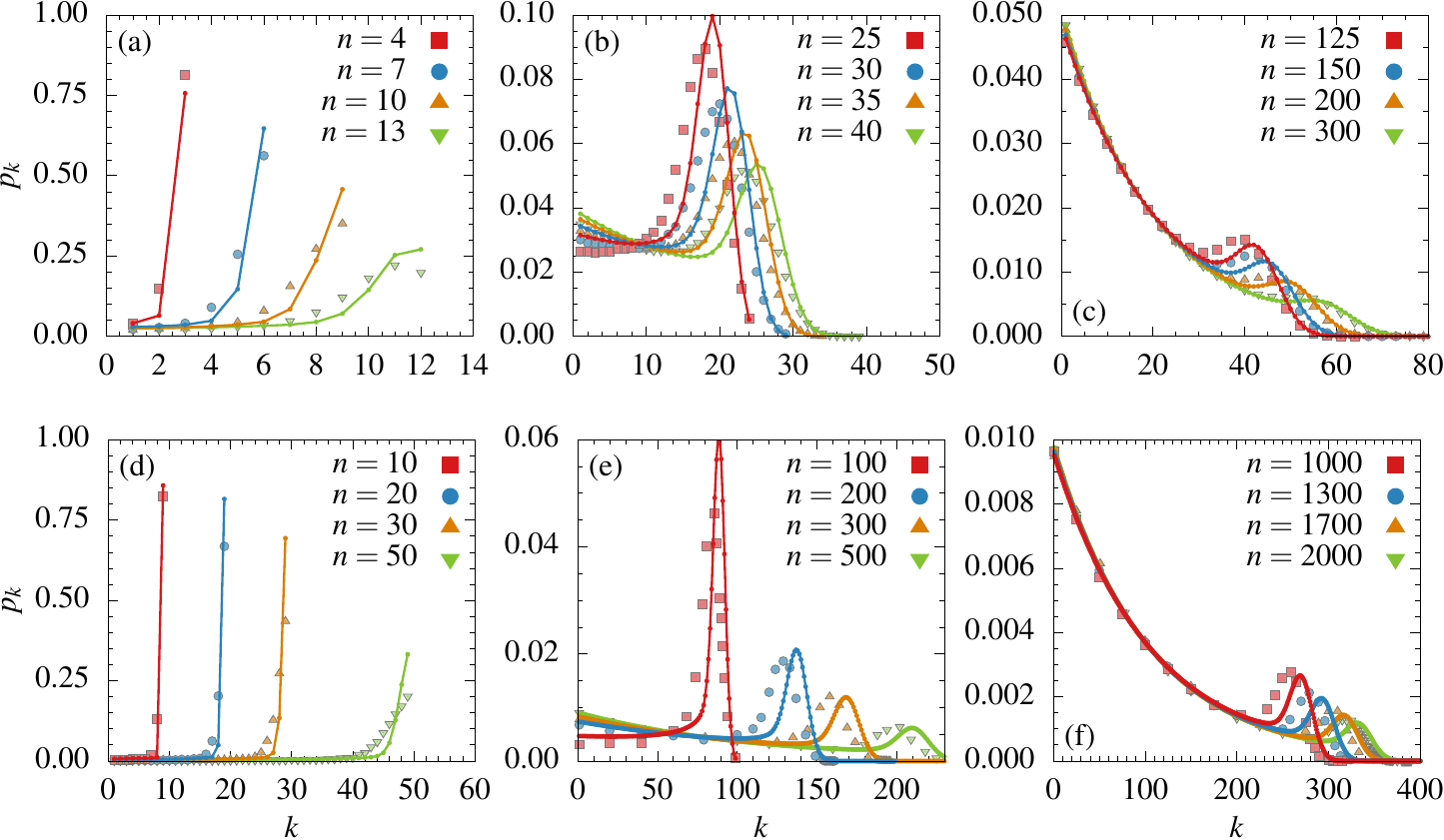}
    \caption{(color online)
        Degree distributions $\{p_k\}$ for various community sizes $n$, with relative event ratios $r = \rho_\ell/\rho_r = 9$ [(a-c)] and $r=49$ [(d-f)]. 
        We compare the solutions of Eq.~\eqref{eq:master_equations_internal_by_size} (small dots) with the average of $20\ 000$ Monte-Carlo simulations (closed symbols).
        Lines are added to the analytical results to guide the eye.
        The analytical expressions are integrated from the initial condition $\vec{p}(t_0)=(p_0=0, p_1=2/3, p_2=1/3)$ with $n=3$ (a dyad), and each distribution (empirical and analytical) corresponds to a snapshot of the average internal degree distribution as the community reaches a fixed average size $n$.
        A bulge appears in the distributions for $n\gg 1$. It is the signature of a \emph{peloton dynamics} \cite{HebertDufresne2012}. We gather some remarks on this dynamics in Appendix~\ref{appendix:peloton}.
    }
    \label{fig:internal_degree_distribution} 
\end{figure*}
% - - - - - - - - - - - - - - - - - - - - - - - - - - - - - - - - - - - - - - -

Summing Eqs.~\eqref{eq:compartemental_master_equation}--\eqref{eq:compartemental_master_equation_n_n}, one finds
\begin{equation}
    \dot{n}(t) \equiv \sum_{k=1}^{\lfloor\!n\!\rfloor} \dot{m}_k (t) =  n\rho_r \; . \label{eq:compartemental_master_equation_n}
\end{equation}
This last equation, together with the observation that $d p_k(t) /dt =  d [m_k(t) / n(t)] /dt$ allows us to describe the system in terms of the average community size $n$ rather than as a function of time.
We find
\begin{align}
    \label{eq:master_equations_internal_by_size}
    \frac{d}{dn}p_k(n) &= \frac{\bigl(p_{k-1} - p_k\bigr)}{n}+r\frac{\bigl(p_{k-1} - p_k\bigr)}{n}  - \frac{p_k}{n}\notag\\
                    &+ \frac{rX}{nZ} \Bigg[(\lfloor\!n\!\rfloor-k)p_{k-1}- (\lfloor\!n\!\rfloor- k - 1)p_{k}\Bigg]\;,
\end{align}
and limit cases similar to the expressions listed in Eqs.~\eqref{eq_group:master_equations_internal}.
This formulation has the added benefit of highlighting the dependency in the \emph{relative} ratio of events $r:=\rho_\ell / \rho_r$.
We validate Eq.~\eqref{eq:master_equations_internal_by_size} in Fig.~\ref{fig:internal_degree_distribution}, where we show that the numerical solutions of this system of differential equations capture the important features of the growth dynamics \footnote{A Python implementation of the integrator is available online at https://github.com/spa-networks/spa.}.
Agreement is, however, not perfect.
Discrepancies between simulations and the solutions of Eq.~(\ref{eq:master_equations_internal_by_size}) can be traced back to the continuous approximation involved in writing differential mean-field equations for discrete quantities, as well as the absence of structural correlation in this type of model.
The net effect is a shift of the prediction toward higher degrees for the bulk of the distribution.

Figure~\ref{fig:internal_degree_distribution} shows  that small and medium communities are highly homogeneous, while the degree distributions in larger communities are heavily skewed.
This heterogeneity arises from the history of the community; the few nodes that join early, when growth is slower, can create more links than the many nodes who join the community as growth accelerates.
The separation in three regimes holds for arbitrary values of $r$, with the transition from homogeneous to heterogeneous degree distributions occurring at higher community sizes $n$ for larger values of $r$ (see the scaling arguments in Appendix \ref{appendix:peloton}).

%~~~~~~~~~~~~~~~~~~~~~~~~~~~~~~~~~~~~~~~~~~~~~~~~~~~~~~~~~~~~~~~~~~~~~~~~~~~~~~
\subsection{Approximate average degree}
\label{subsec:internal_model-mean_field_approximation}
%~~~~~~~~~~~~~~~~~~~~~~~~~~~~~~~~~~~~~~~~~~~~~~~~~~~~~~~~~~~~~~~~~~~~~~~~~~~~~~
A simpler point of view can be adopted to gain further insights into the relation  between the  average degree $\langle k \rangle = 2L/n$ of a node and the size $n(t)$ of its community [$L(t)$ is the number of links in the community at time $t$].

As previously stated, a node will not initiate the creation of new links if its degree equals $n-1$ 
(see Sec.~\ref{subsec:internal_model-description_mean_field}), while the rest of the nodes create new links at a rate $\rho_\ell$.
The total link creation rate is therefore given by 
\begin{equation}
   \frac{dL(t)}{dt} = n(t)\rho_r + n(t)\rho_\ell\bigl[1-p_{n-1}(t)\bigr] \; ,
    \label{eq:approximative_mean_field_L_of_t}
\end{equation}
where $n(t)\rho_r $ is the contribution of the node recruiting process, and where the second term merely states that only nodes of degree $k<n-1$ contribute to the creation of new links within the community at a rate $\rho_\ell$.

If we assume a uniform and uncorrelated distribution of links among nodes, and define $L_{\max}(n)=n(n-1)/2$---the maximal number of links in a community of size $n(t)$---then $p_{n-1}(n)$, the probability that a randomly selected node is of maximal degree $n-1$, can be approximated by
\begin{equation}
    \tilde{p}_{n-1}(n) \simeq \left(\frac{L(n)}{L_{\max}(n)}\right)^{n-1}\;.\label{eq:approx_max_prob}
\end{equation}
Using Eqs.~\eqref{eq:compartemental_master_equation_n} and \eqref{eq:approx_max_prob}, we express the rate of change of $L$ as a function of the average size $n(t)$ at time $t$:
\begin{equation}
    \frac{dL(n)}{dn} = \frac{dL}{dt}\frac{dt}{dn} = 1 + r \left[1-\left(\frac{L(n)}{L_{\max}(n)}\right)^{n-1}\right] \; .
    \label{eq:ER_link_creation_rate}
\end{equation}
While the actual link distribution is neither uniform nor uncorrelated in the model (see Fig.~\ref{fig:internal_degree_distribution}), we will see that our approximation is robust enough, and that Eq.~(\ref{eq:ER_link_creation_rate}) accurately reproduces the average degree (see Sec.~\ref{subsec:internal_model-dunbars_number}).

A simple analysis of Eq.~(\ref{eq:ER_link_creation_rate}) highlights an interesting feature of the model.
For large sizes $n$, the factor $[L(n)/L_{\max}(n)]^{n-1}$ goes rapidly to zero, such that a maximal link creation rate
\begin{equation}
    \frac{dL(n)}{dn} \simeq 1 + r
\end{equation}
is attained.
Hence, the intensive quantity $L(n)/n \to \left(1 + r\right)$ converges toward a constant that depends on the parametrization of the model alone.
Considering that one link equals two stubs (or degree), the asymptotic average degree is directly related to the parameter $r$ through:
\begin{equation}
    \langle k  \rangle = \frac{2L(n)}{n}  \rightarrow 2\left(1+r\right) \; .
    \label{eq:degree_plateau}
\end{equation}
This indicates a maximal average number of connections in a social group.

%~~~~~~~~~~~~~~~~~~~~~~~~~~~~~~~~~~~~~~~~~~~~~~~~~~~~~~~~~~~~~~~~~~~~~~~~~~~~~~
\subsection{Relation with Dunbar's number}
\label{subsec:internal_model-dunbars_number}
\begin{figure}
    \includegraphics[scale=1]{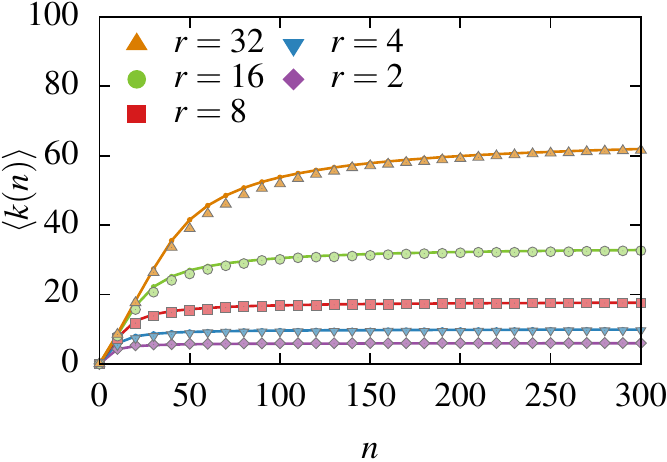}
    \caption{(color online)
        Degree $\langle k (n) \rangle $ of an average node as a function of community size $n$.
        We compare the prediction of Eq.~\ref{eq:ER_link_creation_rate} (lines) with the average of $50\ 000$ Monte-Carlo simulations, for $r=2$ (lozenge), $r=4$ (inverted triangles), $r=8$ (squares), $r=16$ (circle), and $r=32$ (triangles).
    }
    \label{fig:internal_model_mean_avg_degree}
\end{figure}

%~~~~~~~~~~~~~~~~~~~~~~~~~~~~~~~~~~~~~~~~~~~~~~~~~~~~~~~~~~~~~~~~~~~~~~~~~~~~~~
The results shown in Fig.~\ref{fig:internal_model_mean_avg_degree} highlight two different behaviors of the average number of links per individual in relation to the size of a social group.
For low average sizes $n$, the mean degree $\langle k (n) \rangle $ scales linearly with the community size $n$.
In other words, our model captures the fact that \emph{everybody knows everybody within small groups} (e.g., family or close friends).
At larger sizes $n$, $\langle k (n) \rangle $ reaches the plateau $2\left(1+r\right)$ given by Eq.~(\ref{eq:degree_plateau}).
From this point onwards, an average individual will not gain new connections when the potential number of connections is increased.
So, while there is no maximal community size per se, there is a \emph{maximal number of connections that an average individual might possess within a given group} (e.g., large companies or online communities).

% - - - - - - - - - - - - - - - - - - - - - - - - - - - - - - - - - - - - - - -
% Average degree for multiple values of r: Validation of the the plateau equation and mean field description for <k(n)>
% - - - - - - - - - - - - - - - - - - - - - - - - - - - - - - - - - - - - - - -

Interestingly, this upper-bound on the average activity of an individual $\langle k (n) \rangle$ is related to an anthropological theory known as Dunbar's number \cite{Dunbar1992}.
This theory is based on the observed relation between neocortical size in primates and the average size of their social groups.
Its interpretation usually involves information constraints related to the quality of interpersonal connections, and their ability to maintain such relationships.
While the importance of neocortical sizes \cite{DeRuiter2011} and the generality of the results \cite{Shultz2007} are both disputable, the fact remains that empirical evidence supports the existence of an upper bound in the absolute number of active relationships for an average individual, in a given activity (e.g., Ref. \cite{Goncalves2011} for activities on Twitter). In fact, more recent work on social network sizes in humans focus on the progressively higher bounds on average internal degree observed at different social levels or activities: e.g., neighbors, relatives, workplace, and friend circles \cite{hill2003social, dunbar2012social}. These different social levels can be modeled as different communities around one individual (this is the subject of the next section).

In our model, this upper bound naturally emerges and is solely dependent on the parameter  $r$.
This parameter can be interpreted as the ratio between the involvement of an individual in a community, in the sense of bonding with other members, and its contribution to the growth rate of the community.
Note that we do not interpret the plateau as an \emph{absolute} upper bound, but rather as a bound on the maximal number of connections that an \emph{average} individual can maintain.
For low $r$ (or large communities), the rate of change in the population is higher than an individual's involvement such that the maximal degree stagnates.
Whereas, for high $r$ (or small communities), the individual is able to follow the population changes and hence create relationships with most of its members. 
Different types of social organizations will feature different $r$ and, consequently, different values of ``Dunbar's number'' (an online social network, where relationships are easily maintained, will entail higher values of $r$ than a coauthorship network for example):
Different type of activities (networks) should also be modeled using different values of $r$.

In this interpretation, the upper bound on the degree is due to the fact that connections and introduction of new members have linear requirements for individuals, but exponential consequences for the group. 
Other mathematical models describe Dunbar's number (e.g., Ref.~\cite{Goncalves2011}), usually with arguments of priority and/or time and resources management \cite{Dunbar2007}. 
However, our model is based on the \emph{observed} structure of the communities of real networks and consequently, parsimoniously explains Dunbar's number in terms of its two basic units---individuals and groups---and the ratio of their respective characteristic growth rates.
The consequence of this result for the complete community structure of social networks is discussed in Sec. \ref{subsec:dunbar_bis}.
Beforehand, we must first move from a description of the evolution of a single community to a description of the evolution of a superposition of many communities.

%==============================================================================
\section{A growth model for networks with both inter- and intracommunity structure}
\label{sec:complete_model}
%==============================================================================
The model of the previous section is concerned with the growth of an isolated community---a group of friends, a company, or a nascent research group.
Most complex networks, however, comprise more than a single overlapping community \cite{Xie2013}.
To use the model of Sec.~\ref{sec:internal_model} on a larger scale, one therefore needs a mechanism to track multiple, concurrently growing, and overlapping communities.
As we will see shortly, the \emph{structural preferential attachment} (SPA) model of Refs.~\cite{HebertDufresne2011} and \cite{HebertDufresne2012} is both a suitable and practical candidate.

In a nutshell, SPA builds on the popular idea that networks can be interpreted as the projections of abstract structures such as communities \cite{HebertDufresne2015}.
The network is not modeled explicitly: instead, SPA generates an assignment of nodes to overlapping communities, and one instantiates a network based on the community assignments, e.g., by assuming that communities are ER graphs.
SPA therefore lacks an explicit growth mechanism for links.

In what follows, we show how to use the community assignments of SPA jointly with the community growth process of Sec.~\ref{sec:internal_model}.
Specifically, we construct a model in which the history of each community is described by the model of Sec.~\ref{sec:internal_model}, and the history of the community structure is described by SPA.
In this growth model, both facets of the systems---the internal structure and the community structure---evolve simultaneously.
But before we introduce the coupled growth model (in Sec.~\ref{subsec:complete_model-algo}), we first review the key ideas behind SPA.

%~~~~~~~~~~~~~~~~~~~~~~~~~~~~~~~~~~~~~~~~~~~~~~~~~~~~~~~~~~~~~~~~~~~~~~~~~~~~~~
\subsection{Structural preferential attachment}
\label{subsec:complete_model-spa}
%~~~~~~~~~~~~~~~~~~~~~~~~~~~~~~~~~~~~~~~~~~~~~~~~~~~~~~~~~~~~~~~~~~~~~~~~~~~~~~
The essence of SPA can be summarized as follows \cite{HebertDufresne2011}.
At every discrete time step, a growth event occurs.
An event marks the birth of a new node with probability $q$, and the creation of a new fully connected community of $s$ nodes, with probability $p$. 
When an existing node or community is involved (with complementary probabilities $1-q$ and $1-p$ respectively), it is chosen preferentially to its past activity: A node with $x$ memberships or a community of size $x$ is $x$ times more likely to be chosen than a node (or community) with $1$ membership (or node). 
This process ensures that both the membership and size distributions converge to a power-law distribution in the limit of large system sizes.
The probability $q$ controls how interconnected communities are, the probability $p$ controls the distribution of community sizes, and the basic size $s$ allows one to enforce minimal connectivity in the full system.
In SPA, links can only exist between nodes belonging to the same community, and a large-scale connectivity of the network is achieved through overlapping node assignments.

We can write rate equations to follow the numbers $N_u$ of nodes belonging to $u$ groups and the number $S_v$ of groups with $v$ nodes.
These equations are similar to most linear preferential attachment equations
\begin{align}
    N_u(t+&1) = N_u(t) + q\delta_{u,1} \\ 
    & + \frac{1-q+p(s-1)}{t\left[1+p(s-1)\right]}\left[(u-1)N_{u-1}(t) - uN_u(t)\right] \notag\\
     S_v(t+&1) = S_v(t) + p\delta_{v,s} \\ 
    & + \frac{1-p}{t\left[1+p(s-1)\right]}\left[(v-1)S_{v-1}(t)-vS_v(t)\right] \notag \;  ,
\end{align}
and  $N_u$, $S_v$  can be shown to scale as power laws, i.e., $N_u \sim u^{-\gamma_N}$ and $S_v \sim v^{-\gamma_S}$, with exponents \cite{HebertDufresne2012}
\begin{align}
    \gamma_N & = \frac{2-q+2p(s-1)}{1-q+p(s-1)} \;  , \label{eq:gammaN}\\
    \gamma_S & = \frac{2-p + p(s-1)}{1-p} \;  . \label{eq:gammaS}
\end{align}
Because the growth rules are time independent and since $p$ and $q$ are probabilities in the $[0,1]$ interval, the average number of memberships per node and average community size converge in time. Therefore, $\gamma_N$ and $\gamma_S$ are always $\geq 2$ and SPA is not expected to reproduce distributions whose asymptotic decay exponent is smaller than 2.
The interested reader is directed to Refs.~\cite{HebertDufresne2011,HebertDufresne2012} for a complete derivation of these results.

%~~~~~~~~~~~~~~~~~~~~~~~~~~~~~~~~~~~~~~~~~~~~~~~~~~~~~~~~~~~~~~~~~~~~~~~~~~~~~~
\subsection{Coupling a discrete and a continuous processes}
\label{subsec:complete_model-coupling}
%~~~~~~~~~~~~~~~~~~~~~~~~~~~~~~~~~~~~~~~~~~~~~~~~~~~~~~~~~~~~~~~~~~~~~~~~~~~~~~
Recall that our goal is to couple the mechanism of Sec.~\ref{sec:internal_model} (hereafter the \emph{local} model) and SPA.
To do so, we must first determine the relation between the time scales of the local model and that of  SPA, thereby allowing a  concurrent simulation of both processes.
This is not a simple matter, since one must reconcile the continuous nature of the local growth mechanism with the discrete nature of SPA.

In SPA, time $\tilde{T}$ is measured in number of events.
Without loss of generality and for reasons that will become apparent shortly, let us define a rescaled discrete time scale $T$ in which a fraction $\epsilon$ of the time steps lead to SPA events, such that $\epsilon T=\tilde{T}$.
The community structure does not change during the remaining $(1-\epsilon)T$ time steps.
Because a time step $\tilde{T}$ marks the birth of a new community (of size $s$) with probability $p$, or the growth of an existing one with complementary probability $1-p$, we can write the time dependent sum of the sizes $n_i(T)$ of all communities as
\begin{equation}
    \sum_i n_i(T) = \epsilon\ T\bigl[ p s  + (1-p) \bigr] = \epsilon\ T \bigl[1+p(s-1)\bigr].
    \label{eq:sum_of_all_sizes}
\end{equation}
The average size $n_i(T)$ of community $i$ in \emph{discrete} time $T$ is then governed by a rate equation
\begin{align}
  n_i(T+1) &= n_i(T) + \epsilon(1-p) \frac{n_i(T)}{\epsilon\ T \bigl[1+p(s-1)\bigr]}\notag\\
         &= n_i(T)\left[1 + \frac{1}{T}\alpha(p,s)\right]\label{eq:rate_equation_n_discrete_time}
\end{align}
where we have defined $\alpha(p,s):=(1-p)/[1+p(s-1)]$.
Equation \eqref{eq:rate_equation_n_discrete_time} merely states that growth events affect community $i$ with probability $n_i(T)/\sum_j n_j(T)$ (i.e. preferentially to its size).
In the limit of large $T$, \eqref{eq:rate_equation_n_discrete_time} is equivalent to
\begin{equation}
  \frac{dn_i(T)}{dT} = \frac{n_i(T)}{T}\alpha(p,s) \; .
  \label{eq:master_equation_n_discrete_time}
\end{equation}

Now, recall that the size $n_i(t)$ of a community grows exponentially in \emph{continuous} time $t$ as (see Sec.~\ref{sec:internal_model})
\begin{equation}
  \frac{dn_i(t)}{dt} = n_i(t)\rho_r \; .
  \label{eq:master_equation_n}
\end{equation}
Combining the time derivatives \eqref{eq:master_equation_n_discrete_time} and \eqref{eq:master_equation_n}, we obtain a relation between the continuous time $t$ and the discrete time $T$
\begin{equation}
  \frac{dt}{dT} = \frac{dt}{dn_i(t)}\frac{dn_i(T)}{dT} = \frac{n_i(T)}{n_i(t)}\frac{\alpha(p,s)}{\rho_rT} = \frac{\alpha(p,s)}{\rho_rT} \; .
  \label{eq:time_scale_relation}
\end{equation}
% This result holds provided that $n_i(t)=n_i(T)$ at all time.

%==============================================================================
\subsection{The coupled growth model: SPA+}
\label{subsec:complete_model-algo}
%==============================================================================
Equation \eqref{eq:time_scale_relation} tells us how fast a community evolves in comparison with the community structure; we can use this information to formulate an algorithm that simulates both processes concurrently.
We choose to describe the local link creation process of Sec.~\ref{sec:internal_model} in time $T$.
As such, the backbone of the algorithm will be the SPA process, to which we now must add details pertaining to the local model of Sec.~\ref{sec:internal_model}.

The first part of the local model (nodes are recruited at rate $n\rho_r$) is easily accounted for: Whenever a node joins a new community, we simply choose a recruiting node uniformly among the current members of that community and form a new link.
The exponential growth of communities in SPA ensures that this process is consistent with the model of Sec.~\ref{sec:internal_model}.

The second part of the local model (links are created at rate $\propto n\rho_\ell$) entails a more involved analysis.
Let us define $\tilde{n}_i(t)$ as the \emph{effective} size of community $i$, i.e., the number of nodes that are allowed to create links [number of nodes of degree $k < n_i(t)-1$ links within community $i$].
Then, in the local model, the number of links $L(t)$ in a community of effective size $\tilde{n}_i(t)$ grows at a rate
\begin{equation}
    \frac{dL(t)}{dt} = \rho_\ell\ \!\tilde{n}_i(t)
\end{equation}
such that links are introduced in the community at the rate
\begin{align}
    \frac{dL(T)}{dT} & = \frac{dL(t)}{dt}\frac{dt}{dT} = \frac{\rho_\ell}{\rho_r} \frac{\tilde{n}_i(t)}{T} \alpha(p,s)\notag\\
                     &=  r\epsilon(1-p)\frac{\tilde{n}_i(T)}{\sum_j n_j(T)} \; . 
                     \label{eq:internal_link_creation_rate_discrete}
\end{align}
The purpose of time transformation $\epsilon\ T = \tilde{T}$ is then apparent: It can be adjusted to bound $r\epsilon(1-p)$ to the interval $[0,1]$  for all $r\in\mathbb{R}^+$.
Since $\epsilon$ is an arbitrary fraction which also lies in $[0,1]$, we adopt the simplest choice, i.e.,
\begin{equation}
    \epsilon =
    \begin{cases}
        [r(1-p)]^{-1} & \text{if $r(1-p)>1$},\\
        1 & \text{otherwise.}
    \end{cases}
    \label{eq:epsilon}
\end{equation}
Equation \eqref{eq:internal_link_creation_rate_discrete} can then be interpreted in two ways.
Straightforwardly, we may say that at each time step $dT$ of the SPA process, a new link is created between the existing members of a community of effective size $\tilde{n}_i$ with a probability given by the right-hand side of \eqref{eq:internal_link_creation_rate_discrete} for all $i$.
Alternatively, we may say that at each time step $dT$ of the SPA process, a new link is created with probability $r\epsilon\left(1-p\right)$ in a community selected with a probability proportional to its effective size $\tilde{n}_i(T)/\sum_j n_j(T)$.
Equation \eqref{eq:epsilon} ensures that this interpretation is always sensible.
In both interpretations, if a link must be created, we choose two nodes of degree $k<n_i(T) - 1$ at random and connect them.

Note that the ratio $\tilde{n}_i(T)/\sum_j n_j(T)$ is not normalized.
In the context of the second interpretation, this implies that at each time step $dT$, there is a probability $1 - \sum_i \tilde{n}_i(T)/\sum_j n_j(T)$ that no link creation event will  occur.
Alternatively, we may select the community in which the link creation event occurs proportionally to its \emph{actual} size $n_i(T)$ and connect two nodes chosen uniformly among  \emph{all} the nodes of that community.
The ratio $\tilde{n}_i(T)/\sum_j n_j(T)$ will then be effectively respected if we consider that a link creation simply ``fails'' whenever the first randomly selected node has the maximal number of connections.\\

The above analysis yields a straightforward algorithm for the modified version of SPA (hereafter SPA+)~\footnote{A C++11 implementation of SPA+ is available online at https://github.com/spa-networks/spa.}.
Starting with disjoint and fully connected communities of size $s$, at each discrete time step $T$:
\begin{itemize}
    \item[\textbf{1}:] a new community of size $s$ is created with probability $p\epsilon$ or an existing one (chosen preferentially with respect to its size) grows with probability $(1-p)\epsilon$;
    \begin{itemize}
        \item[\textbf{1.a}:] if a community birth event occurs, one of the $s$ involved nodes is a new one with probability $q$ or an existing one (chosen preferentially with respect to its current number of memberships) with complementary probability $1-q$.
        The other $s-1$ nodes are chosen preferentially with respect to their current number of memberships among existing nodes;
        \item[\textbf{1.b}:] if a community growth event occurs, the involved node is a new one with probability $q$ or an existing one (chosen preferentially with respect to its current number of  memberships) with probability $1-q$.
        Once the node is added to the community, we randomly select another node in the community (uniformly) and create a link;
    \end{itemize}
    \item[\textbf{2}:] with probability $r\left(1-p\right)\epsilon$, a new link is created in a community chosen preferentially to its size.
    It connects a uniformly chosen node, and a uniformly chosen potential neighbor, provided that the source node is not already connected to every node in the community.
\end{itemize}
If  $r(1-p) < 1$, link creation occurs on slower time scale than community structure related events, whereas the converse is true if $r(1-p)>1$.

% - - - - - - - - - - - - - - - - - - - - - - - - - - - - - - - - - - - - - - -
% Note: This figure appears long before the text that references it; this pushes
% the figure earlier in the PDF, and closer to associated text. 
% - - - - - - - - - - - - - - - - - - - - - - - - - - - - - - - - - - - - - - -
% Global statistics of empirical networks
\begin{figure*}[ht]
    \includegraphics[scale=1]{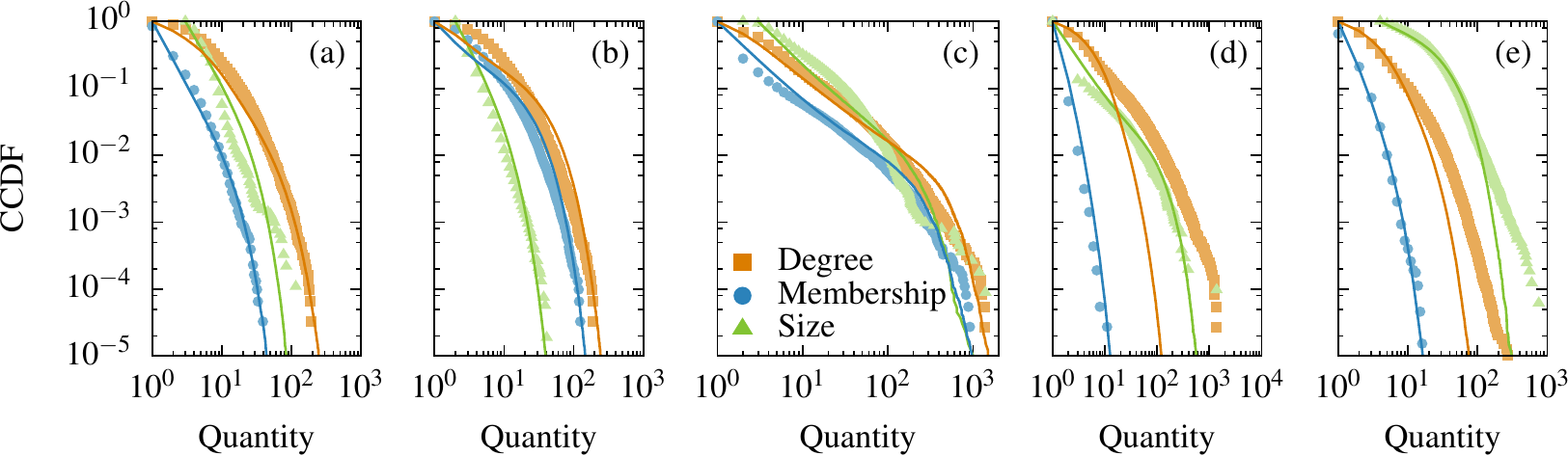}
    \caption{(color online)
        Reproduction of the global statistics of the community structure of real networks by SPA+.
        We detect the community structure with standard algorithms, and then fit SPA+ to the global statistics of the resulting network--overlapping community structure pair (see Appendix~\ref{appendix:parameters} for details).
        The networks, community detection algorithms and parameters (in square brackets) are the following: (a) aXiv with CCPA [$0.60, 0.58, 4.74$] (b) arXiv with LCA [$0.61, 0.16, 0.76$] (c) Enron with LG [$0.25, 0.18, 1.59$] (d) Enrom with OSLOM [$0.20 , 0.85, 2.2$] and (e) MathSciNet with GCE [$0.07, 0.75, 0.74$].
        Empirical complementary cumulative distribution functions (CCDFs) are shown using closed symbols.
        The solid curves are obtained by averaging the corresponding quantities over 200 realizations of SPA+.
        The initial condition of each simulation contains between $N/100$ and $N/10$ disconnected nodes, and the simulation is stopped when the network reaches its final sizes of $N=30\ 561$ nodes for arXiv 
        [(a) and (b)], $N=36\ 692$ nodes for Enron [(c) and (d)], and $N=391\ 529$ nodes for MathSci (e).
        All reproductions are realized with $s=1$.
    }
    \label{fig:reproduction_real_network}
\end{figure*}
% - - - - - - - - - - - - - - - - - - - - - - - - - - - - - - - - - - - - - - -
%==============================================================================
\subsection{Redundant memberships, multiple links and self-loops}
\label{subsec:complete_model-finite}
%==============================================================================
In SPA, one assumes that a community grows on its own, and that new members are drawn from an infinite reservoir of indistinguishable nodes \cite{HebertDufresne2012}.
In practice, the reservoir is finite and each node therein is tagged; when the system is small and the parameters $p$ and $q$ take extreme values ($q \simeq 0$ for any $p$, the worst case being $q \simeq 0$ and $p \simeq 0$), there is a significant probability that a node will appear more than once in a community.
To respect the relative rates of all events and preserve the mean-field mapping of Sec~\ref{subsec:complete_model-coupling}, we consider that these duplicate nodes are effectively new.
The implications of this observation for the community structure are discussed at length in Ref.~\cite{HebertDufresne2012}.
There is additional implications for the combined SPA+ model.

The fact that the same node can (and will) join the same community more than once implies that we will create parallel links and self-loops, because a node can become connected to copies of itself.
Because these types of links are seldom considered in empirical datasets, we collapse the redundant memberships into a single membership at the end of the growth process, i.e., we merge nodes with all their duplicates within each communities.
This  (a) skews the tail of the membership and size distribution and (b) removes multiple self-loops from the system.
The net effect is that communities becomes denser on average.
We note that these redundant memberships are known to account for a vanishingly small fraction of all memberships when the number of communities is large and the $(p,q)$ parameters are not too small \cite{HebertDufresne2012}.
The consequences of redundant memberships should therefore subside in large networks.

%==============================================================================
\section{Results and Discussion}
\label{sec:results}
%==============================================================================
The SPA model has previously been shown to capture many properties of the community structure of real networks \cite{HebertDufresne2011,HebertDufresne2012}, such as the distribution of community sizes, of node memberships, and of community degrees.
We now investigate these properties anew by modeling three social networks: Two coauthorship networks obtained from the arXiv circa 2005 \cite{Palla2005} and from MathSciNet circa 2008 \cite{Palla2008}, as well as the email exchange network of Enron \cite{Klimt2004}.
We detect their community structure with five different algorithms: A link clustering algorithm \cite{Ahn2010} (LCA), a greedy clique expansion algorithm \cite{Lee2010} (GCE), the order statistics local optimization method \cite{Lancichinetti2011} (OSLOM), a greedy modularity optimization of line-graphs algorithm \cite{Evans2009} (LG), and a modified version of the classical clique percolation algorithm \cite{Young2015,Palla2005} (CCPA).
This provides us with a total of 15 systems, from which we have selected 5 representative examples: arXiv as described by both the CCPA and LCA, Enron as described by the GCE and OSLOM algorithms, and MathSciNet as described by the GCE algorithm.
Note that three of the above algorithms (LCA, LG, CCPA) identify \emph{link} partitions, while the other two directly find overlapping \emph{node} communities.
We translate link partitions into node communities to analyze every algorithm on a common basis, where the true community of a link is unknown.

We model a real network by estimating a value for the tuple of parameters $(\hat{p},\hat{q},\hat{r})$.
The details of the parameter estimation procedure are gathered in Appendix~\ref{appendix:parameters}.
In a nutshell, we use the community structure of the real network to first estimate $p$ and $q$ (yielding $\hat{p}$ and $\hat{q}$). 
We then obtain an estimate $\hat{r}$ of $r$ by fitting the model of Sec~\ref{sec:internal_model} to the internal degree distributions of each community.
The final number of nodes $N$ and the basic community size $s$ are both fixed by the empirical dataset. 
$N$ is trivially the number of nodes in the real network, and we select $s=1$ in all cases, because it leads to networks with more than one component, a feature of the empirical datasets listed above.

The SPA+ model is, in some sense, minimal.
One parameter controls the amount of overlap ($q$), one parameter controls the distribution of community sizes ($p$), and one parameter controls the density of these communities ($r$).

%~~~~~~~~~~~~~~~~~~~~~~~~~~~~~~~~~~~~~~~~~~~~~~~~~~~~~~~~~~~~~~~~~~~~~~~~~~~~~~
\subsection{Global statistics}
\label{subsec:modeling_real_networks}
%~~~~~~~~~~~~~~~~~~~~~~~~~~~~~~~~~~~~~~~~~~~~~~~~~~~~~~~~~~~~~~~~~~~~~~~~~~~~~~
In Fig.~\ref{fig:reproduction_real_network} we compare the statistical properties of SPA+ networks with their empirical counterparts.
In this respect, the new contribution of the present study is the global degree distribution: SPA models the distribution of community sizes and node memberships, while the growth mechanism of Sec.~\ref{sec:internal_model} models the degree distribution within each community.
The degree distribution of the network is an emerging property of the SPA+ model, since it is not modeled directly.
It is necessarily fat tailed, because it arises from the convolution of two fat tailed distributions (memberships and sizes) \cite{HebertDufresne2012}.
The parameter $r$ [and thus the local model of Sec.~\ref{sec:internal_model}] controls the speed of the decay of the degree distribution, through its effect on the relation between community size and average degree.

Figure~\ref{fig:reproduction_real_network} shows that SPA+ can reproduce the degree distribution of the real dataset, if the overlapping communities decomposition of the network is in line with our modeling hypotheses.
That is, SPA+ can generate degree distributions with the correct shape only if the detected community structure is heterogeneous.
By heterogeneous, we mean that the distributions of community sizes and node memberships are either power laws, or power laws with an exponential cut-off.
As long as we consider such systems, we can fit both the size and membership distribution robustly \cite{Clauset2009,HebertDufresne2012} 
[see Figs.~\ref{fig:reproduction_real_network}(a)--\ref{fig:reproduction_real_network}(c)]. 
Due to the nature of our model, the quality of the predicted degree distribution is inherently connected to the quality of the predicted size and membership distributions. 
SPA+ does poorly in two cases [see Figs.~\ref{fig:reproduction_real_network}(d)--\ref{fig:reproduction_real_network}(e)], and since the membership distributions are well represented in all cases studied, the culprits lay mainly with the size distributions. 
In Fig.~\ref{fig:reproduction_real_network}(d), we diverge from the data at low community sizes and fail to account for an extremely large community (of size $n=1384$). 
In Fig.~\ref{fig:reproduction_real_network}(e),  the empirical size distribution decays asymptotically slower than the behavior accessible to the model, i.e., $\gamma_N \geq 2$ [Eq.~\eqref{eq:gammaN}].
We also note that the statistics of real datasets do contain kinks and bumps (real or spurious) that cannot be reproduced by simple growth models like SPA+, although the average behavior can be well captured [see Figs.~\ref{fig:reproduction_real_network}(a) and \ref{fig:reproduction_real_network}(c)].

%~~~~~~~~~~~~~~~~~~~~~~~~~~~~~~~~~~~~~~~~~~~~~~~~~~~~~~~~~~~~~~~~~~~~~~~~~~~~~~
\subsection{Local statistics}
%~~~~~~~~~~~~~~~~~~~~~~~~~~~~~~~~~~~~~~~~~~~~~~~~~~~~~~~~~~~~~~~~~~~~~~~~~~~~~~
% - - - - - - - - - - - - - - - - - - - - - - - - - - - - - - - - - - - - - - -
% Internal degree distribution in empirical networks
\begin{figure}
    \centering
    \includegraphics[scale=1]{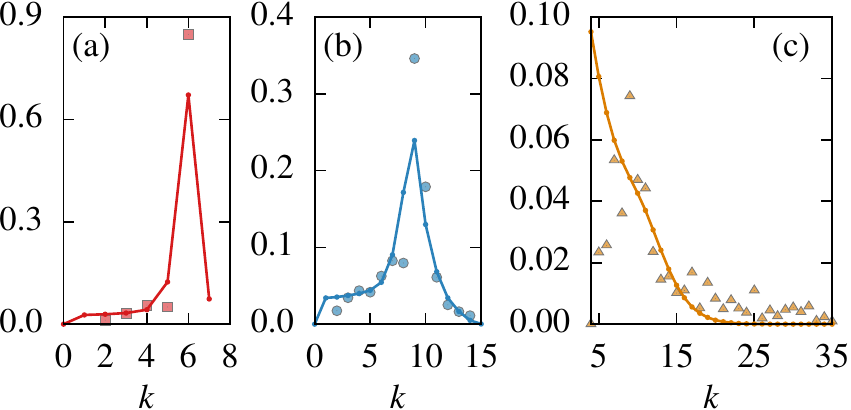}
    \caption{Average internal degree distribution in arXiv with overlapping communities detected by CCPA.
    The results are averaged over communities of sizes (a) $n=7$ [646 communities], (b) $n=10, \ldots ,15$ [340 communities], and (c) $n=30, \ldots ,200$ [26 communities].
    We account for the relative contribution of each community size by plotting the averaged probability $P(k|n\in[n_a, n_b])= \sum_{n=n_a}^{n_b} p_k(n) \omega (n_a) / \Omega_{ab}$, where $\omega(n)$ is the number of communities of size $n$ in the network and $\Omega_{ab}=\sum_{n=n_a}^{n_b} \omega(n_a)$.
    % Note that $P(k|n\in[n_a, n_b])$ is not a distribution.
    % We show the degree distribution within communities of a given size $n$.
    The solid line shows the prediction of the internal model Sec.~\ref{sec:internal_model}), with $r=10$ (a), $r=7$ (b), and $r=2$ (c).
    The analytical prediction is averaged with the weights $\{\omega(n_a)\}$ of the real network--community structure pair.
    The important qualitative features of each regime is captured by the model: The distribution becomes increasingly heterogeneous with growing values of $n$ and the peak of $P(k|n)$ moves towards lower degrees.
    Note that the empirical data cannot peak at $k=1$, 
   since the CCPA detects communities by combining cliques of size $\geq 3$ \cite{Young2015}.}
    \label{figure:empirical_internal}
\end{figure}
% - - - - - - - - - - - - - - - - - - - - - - - - - - - - - - - - - - - - - - -
In Sec.~\ref{sec:internal_model}, we have established that according to our model, the internal degree distributions of growing communities could display three different regimes:
A highly homogeneous regime where every node is nearly of the maximal degree, a homogeneous regime where the bulk of the nodes has similar degrees, and a heterogeneous regime where the majority of the nodes have low degrees (while a few nodes are highly connected).
These regimes can be observed in a number real networks, once their community structure is uncovered by algorithms designed for the detection of overlapping communities. In Fig.~\ref{figure:empirical_internal}, we present the three regimes in the arXiv co-authorship network, as detected by the CCPA algorithm.
The figure illustrates two important facts.
On the one hand, it puts the internal model of Sec.~\ref{sec:internal_model} on firmer empirical ground---it confirms that the evolution of the internal degree  distribution of arXiv is captured by the model.
On the other hand, it emphasizes that the internal degree distributions of the uncovered communities can be quite distinct from random Erd\H{o}s-R\'enyi graphs, as it is often implicitly assumed.
This further supports the recent shift towards principled community detection algorithms which explicitly allow for arbitrary degree distributions within communities \cite{Peixoto2012,Karrer2011,Peixoto2015}.

The results of Fig.~\ref{figure:empirical_internal} must, however, be taken with some caution.
We have not performed an exhaustive search, instead we have selected a network well reproduced by SPA+ (see Fig.~\ref{fig:reproduction_real_network}), and have averaged the distribution not only over all communities of the same size $n$, but also over many community sizes.
This procedure was necessary since there is only a handful of communities at any given size $n\gg 1$.
A more thorough study of real (large) networks will be able to tell us just how prevalent the separation in three regimes actually is.

% - - - - - - - - - - - - - - - - - - - - - - - - - - - - - - - - - - - - - - -
% Note: This figure appears long before the text that references it; this pushes
% the figure earlier in the PDF, and closer to associated text. 
% - - - - - - - - - - - - - - - - - - - - - - - - - - - - - - - - - - - - - - -
% Correlation between memberships, community size and average internal degree
\begin{figure}
    \centering
    \includegraphics{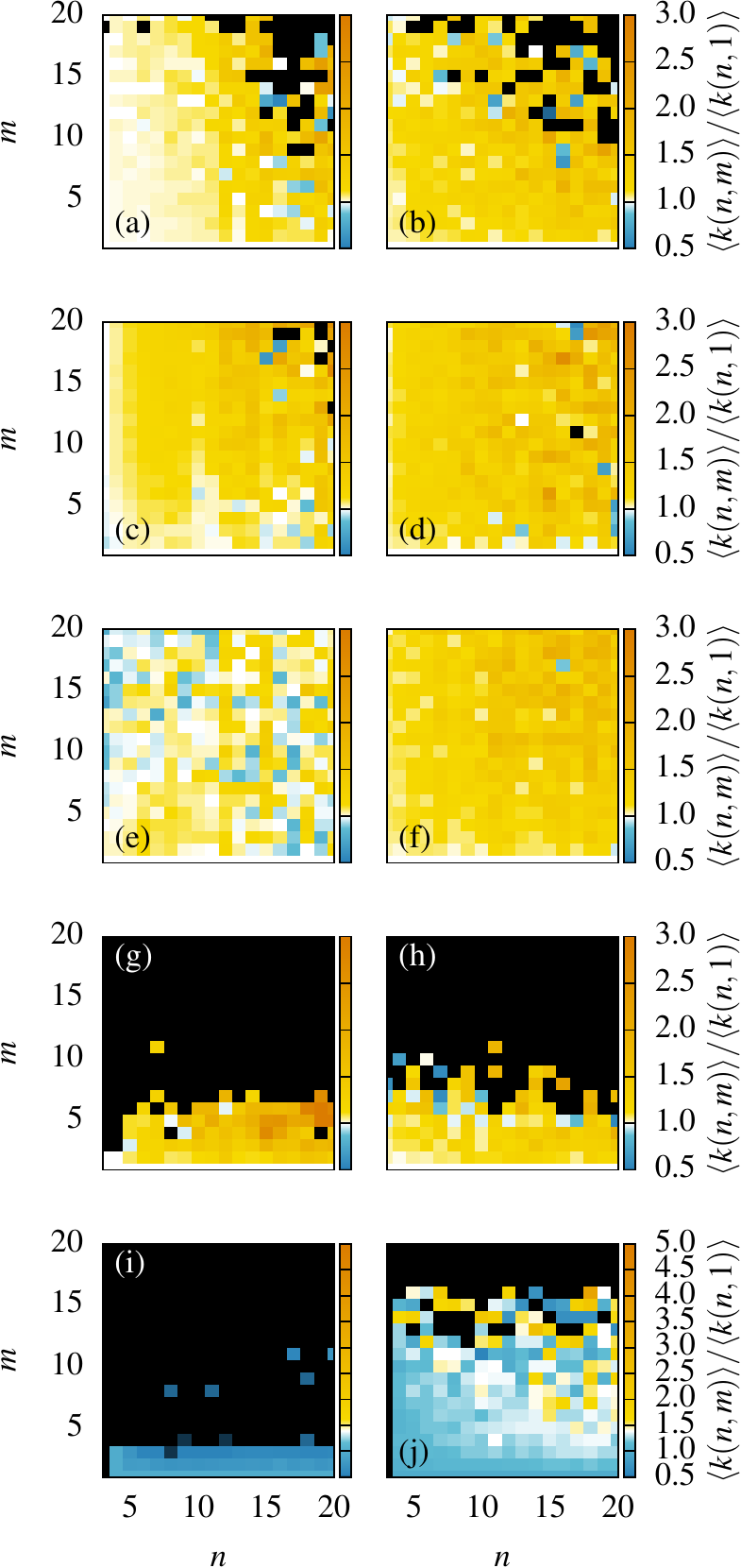}
    \caption{(color online)
        Correlations among node memberships $m$, community sizes $n$ and average internal degree $\langle k(n,m) \rangle$ measured in relation to $\langle k(n,1) \rangle$.
        Results are shown for [(a-d)] arXiv, [(e-h)] Enron, and [(i) and (j)] MathSciNet, using [(a) and (b)] CCPA, [(c) and (d)] LCA, [(e) and (f)] LG, [(g) and (h)] OSLOM, and [(i) and (j)] GCE.
        Real networks appears on the left-hand side [(a), (c), (e), (g), and (i)] and we show a single realization of an equivalent SPA+ network on the right-hand side [(b), (d), (f), (h), and (j)], see the caption of Fig.~\ref{fig:reproduction_real_network} for the parameters.
        Black squares indicate missing data.
        Note that, without the addition of the link creation mechanism of Sec.~\ref{sec:internal_model}, the SPA model does not include any correlations, even when one considers a given density function of community sizes (i.e., in SPA,  $\langle k(n,m) \rangle = \langle k(n,1) \rangle$ on average, for all $n$ and $m$).
    }
    \label{fig:reproduction_real_network-correlations}
\end{figure}
% - - - - - - - - - - - - - - - - - - - - - - - - - - - - - - - - - - - - - - -

% - - - - - - - - - - - - - - - - - - - - - - - - - - - - - - - - - - - - - - -
% Note: This figure appears long before the text that references it; this pushes
% the figure earlier in the PDF, and closer to associated text. 
% - - - - - - - - - - - - - - - - - - - - - - - - - - - - - - - - - - - - - - -
% Global statistics of empirical networks
% Density and average degree in empirical networks
\begin{figure*}
    \centering
    \includegraphics[scale=1]{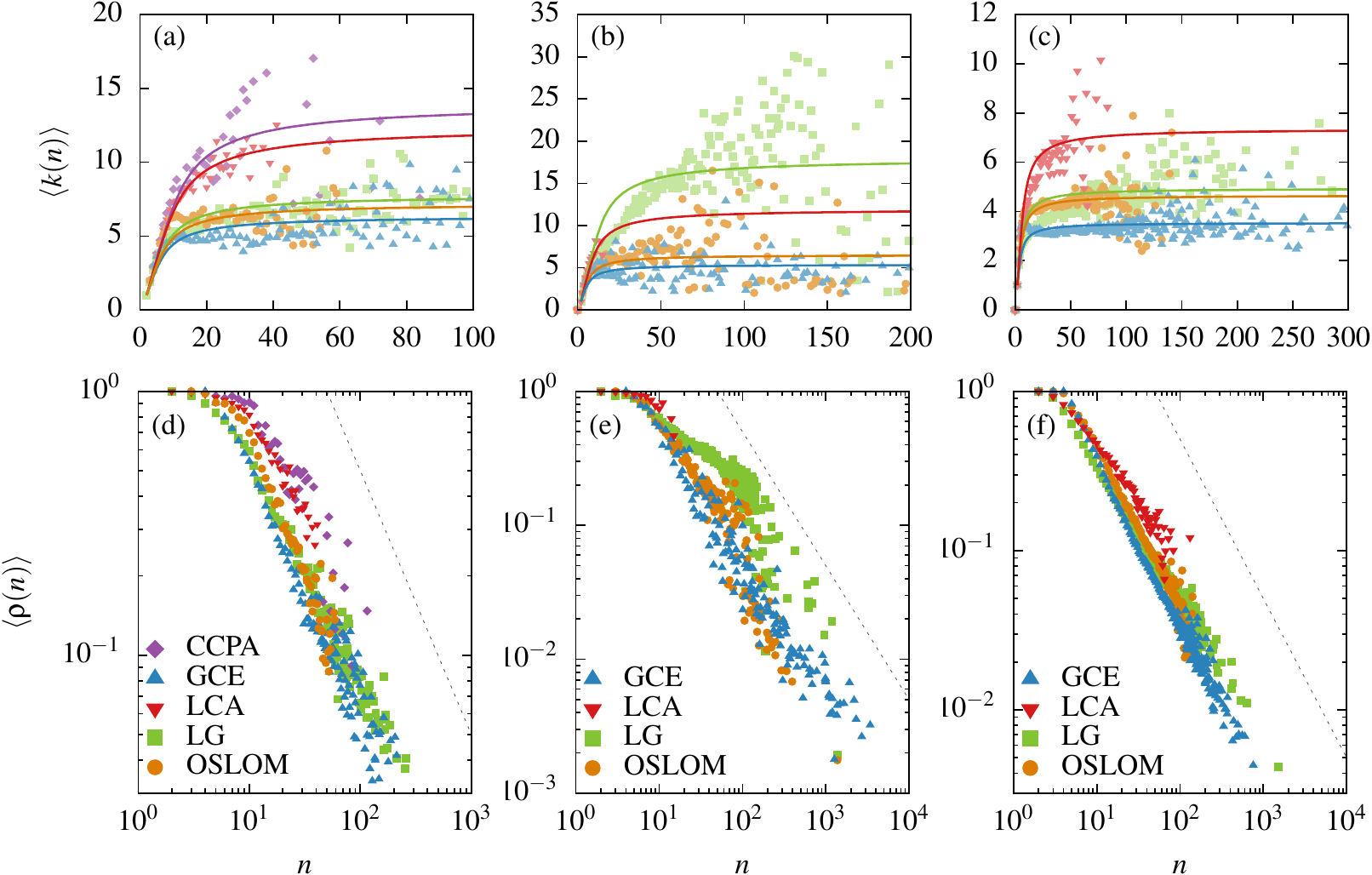}
    \caption{(color online)
        Dunbar's number in empirical data.
        [(a)-(c)] Average number of connections $\langle k (n)\rangle$ for individuals within a community of size $n$.
        [(d)-(f)] Average density $\langle \rho (n)\rangle = \langle k \rangle / (n-1)$ of communities of size $n$.
        The networks are [(a) and (d)] arXiv, [(b) and (e)] Enron, and [(c) and (f)] MathSciNet.
        A line $\langle \rho (n)\rangle \propto n^{-1}$ is traced to guide the eye.
        The uncorrected estimates of $\hat{r}$ (see Appendix \ref{appendix:parameters}.3.d) using the LSE are, from bottom to top, (arXiv) $\hat{r}=2.18, 2.61, 2.89, 5.20, 5.99$ (Enron) $\hat{r}=1.69, 2.27, 4.97, 7.96$ and (MathSci) $\hat{r}=0.77, 1.33, 1.47, 2.68$.
    }
    \label{fig:reproduction_real_network-density}
\end{figure*}
% - - - - - - - - - - - - - - - - - - - - - - - - - - - - - - - - - - - - - - -

%~~~~~~~~~~~~~~~~~~~~~~~~~~~~~~~~~~~~~~~~~~~~~~~~~~~~~~~~~~~~~~~~~~~~~~~~~~~~~~
\subsection{Correlations between the global community structure\\ and the local structure of communities}
%~~~~~~~~~~~~~~~~~~~~~~~~~~~~~~~~~~~~~~~~~~~~~~~~~~~~~~~~~~~~~~~~~~~~~~~~~~~~~~
An additional property is also captured by SPA+.
The results shown in Fig.~\ref{fig:reproduction_real_network-correlations} investigate correlations between the organization within communities and the overarching community structure. 
We obtain the relation between the average internal degree of a node within communities of size $n$ (i.e., the ``social involvement'' of an individual within a group), and its membership number $m$, in empirical datasets and the corresponding simulated  networks.
We quantify this relationship by the ratio  $\langle k(n,m) \rangle / \langle k(n,1)\rangle$.

Generally, all algorithms except GCE find that nodes active in the community structure (high number $m$ of memberships) tend to be also active within communities (high average internal degree $\langle k \rangle$).
Even though agreement is not perfect, our model reproduces this effect through age-memberships and age-degree correlations.
While the available data do not tell whether these correlations are indeed age related, it is natural to assume that authors or employees who have been active for a longer time the arXiv or a company, tend to have both more social groups and more relations within them.
To the best of our knowledge, these correlations are not considered in other growth models, but naturally emerge here, from our link creation mechanism.
In essence, this means that individuals acting as hubs in the community structure (many memberships), tend to act also as hubs within the structure of their communities.

These remarks bear some relation to the hub dichotomy, first introduced in the literature of protein-protein interaction networks \cite{han2004evidence, bertin2007confirmation,agarwal2010revisiting}, namely the distinction between date hubs (nodes with many links in different communities) and party hubs (nodes with many links from a given community).
What we see in social networks is that there also exists a different and important class of hubs with many links from many different communities.
This stresses anew the importance of nodes that act as social bridges by connecting different communities \cite{nepusz2008fuzzy, wang2011identifying}. While these hubs have long been recognized as important \cite{granovetter1973strength}, they are now also a focus of immunization methods on networks \cite{masuda2009immunization, hebert2013global}. 

%~~~~~~~~~~~~~~~~~~~~~~~~~~~~~~~~~~~~~~~~~~~~~~~~~~~~~~~~~~~~~~~~~~~~~~~~~~~~~~
\subsection{Some implications of Dunbar's number}
\label{subsec:dunbar_bis}
%~~~~~~~~~~~~~~~~~~~~~~~~~~~~~~~~~~~~~~~~~~~~~~~~~~~~~~~~~~~~~~~~~~~~~~~~~~~~~~
In Sec.~\ref{subsec:internal_model-dunbars_number}, we have discussed the theoretical relation between our model for the internal structure of communities and a cognitive limit in an individual's social relationships known as Dunbar's number.
In our model, this limit stems from the ratio of effort put into building new connections $\rho_r$ and in increasing group size $\rho_\ell$, which constraints the average internal degree in large groups.
In Fig.~\ref{fig:reproduction_real_network-density}, we observe a similar behavior in our social network datasets.
The empirical results are also compared with our model using the least-squares estimator (see Appendix~\ref{appendix:parameters} for details).

In the context of overlapping communities, we wish to emphasize three important caveats on the connection between our work and recent studies on Dunbar's number. First, most work on bounds of active relationships in different communities is concerned with nested social levels
 \cite{zhou2005discrete, hamilton2007complex}. While our communities overlap, they are not in any way nested. Second, on a related issue, if we wish to interpret different communities as a node's family, friends, or workplace, we should allow nodes to have different involvement $r$ in different communities. Third, if on the other hand we wish to interpret an entire network as one level of activity, Dunbar's number then implies a bound on a node's total degree. While both the internal average degree per community and the number of communities per node are bounded, we have shown strong correlations between these two quantities.
Actually, one can easily infer from the algorithmic description of the model (see Sec.~\ref{subsec:complete_model-algo}) that the average degree converges to $(1-p)(1+r)/q$ and is thus also bounded.

Finally, the observed plateau in internal degree implies a vanishing average density $\langle \rho (n) \rangle $, i.e., fraction of potential links that exist, for large communities.
Regardless of the nature of the network, of the community detection algorithm and of the parameters $(p,q,r)$, the simple existence of the plateau implies that community density vanishes as $\langle \rho (n) \rangle  \sim n^{-1}$.
This is obviously true in our model, and observed for our datasets in Fig.~\ref{fig:reproduction_real_network-density}.
Only the community structure of Enron as detected by LG stands out from the prediction.
Further empirical studies would, however, be required to support this finding.

%==============================================================================
\section{Conclusion}
\label{sec:conclusion}
%==============================================================================
We have introduced a simple model for the growth of a community and focused on its connection with a model that describes the growth of overlapping community structures (SPA).
In so doing, we have showed that the local model is consistent with empirical observations (vanishing density and varying heterogeneity in communities).
We have then explored a number of properties of the combined model (SPA+) and investigated the same properties in empirical networks.
These properties came in three categories: global statistics (distributions of sizes, memberships and degrees), correlations between a node's activities within communities and within the overarching community structure, and the vanishing density (as $\sim n^{-1}$) of large communities.
In all cases, we  have found that SPA+ behaves much like its empirical counterpart.
We have also shown that our model is consistent with the theory of Dunbar's number, both within communities and at the level of the complete network.
The presentation of shortcomings and successes of the SPA+ principle (in terms of predictive value) shows the importance and the need for further study in stochastic growth models.

%==============================================================================
\section*{Acknowledgments}
%==============================================================================
We are grateful to anonymous referees for their suggestions in improving our presentation.
The authors thank Calcul Qu\'ebec for the computing facilities.
This work has been supported by the Instituts de recherche en sant\'e du Canada, the Conseil de recherches en sciences naturelles et en g\'enie du Canada, the Fonds de recherche du Qu\'ebec-Nature et technologies, and the James S. McDonnell Foundation Postdoctoral Fellowship.
J.-G.Y. and L.H.-D. contributed equally to this work.

%==============================================================================
\appendix
\section{Peloton dynamics}
\label{appendix:peloton}
%==============================================================================
% - - - - - - - - - - - - - - - - - - - - - - - - - - - - - - - - - - - - - - -
% Renormalization procedure
\begin{figure}
    \includegraphics{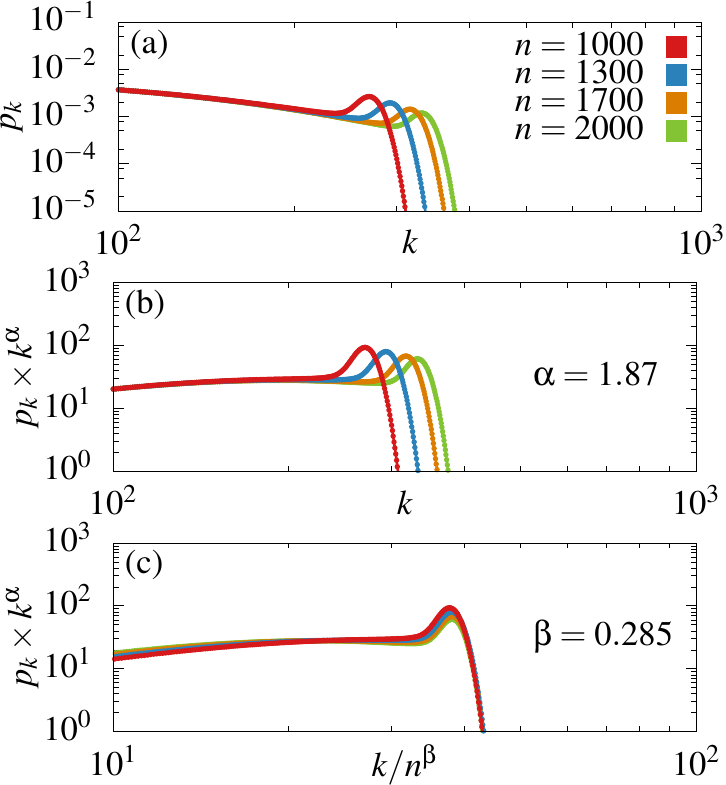}
    \caption{(color online)
        Scaling analysis of the internal degree distributions of the model of Sec.~\ref{sec:internal_model} [numerical solutions of Eq.~\eqref{eq:master_equations_internal_by_size} with $r= 49$].
        (a) The distributions $p_k(n;r)$ of Fig.~\ref{fig:internal_degree_distribution}(f) on a log-log scale for $n= 1000, 1300, 1700, 2000$; 
        (b) the scaled distributions $p_k(n;r) \times k^{\alpha(r)}$ with $\alpha(r) = 1.87$;
        (c) the scaled distributions as a function of the scaled variable $k/n^{\beta(r)}$ with $\beta(r)= 0.285$.}
    \label{fig:peloton}
\end{figure}
% - - - - - - - - - - - - - - - - - - - - - - - - - - - - - - - - - - - - - - -
This Appendix presents our preliminary analysis of the results of Fig.~\ref{fig:internal_degree_distribution} which are reminiscent of the peloton dynamics studied in Ref. \cite{HebertDufresne2012}.
It is a finite-size effect related to the leaders dynamics; groups of highly connected individuals result in a clearly identifiable bulge in the degree distribution. 
Averaging over multiple realizations of the growth of a community leads to the creation of a peloton where one is significantly more likely to find entities than predicted by the asymptotic distribution.
Because the same peloton evolves with growing $n$, it is expected to retain its shape across a large range of  community sizes.
The simplest scaling ansatz takes the form
\begin{align}
p_k(n;r) \simeq k^{-\alpha(r)}\  G(k/n^{\beta(r)}) \qquad {\rm for} \qquad n \gg 1\;, \label{eq:Ansatz} 
\end{align}
where $G(x)$ is a universal function.
The construction is clear: $k^{-\alpha}$ takes care of the power law decreases and the scaled variable $k/n^\beta$ aligns all curves together.
This exercise is carried out in Fig.~\ref{fig:peloton} for the case $r= 49$.
The procedure is inspired by Ref.~\cite{Christensen2005} and is called quite appropriately \textit{data collapse}.

Although, we have not investigated the exact form of $G(x)$, its general behavior is characteristic of a number of self-organized critical systems observed thus far (Ref.~\cite{Christensen2005}): A flat curve sharply rising to a well defined maximum followed by a rapid exponential decrease as a function of the rescaled variable.
The scaling information is captured by the exponents $\alpha$ and $\beta$. 
They can be extracted numerically from the positions $k_b(n; r)$ of the maxima of the bulges of the individual probability distributions, together with the values of the probabilities $p_{k_b}$ at these maxima [see Fig.~\ref{fig:peloton}(a)] and the scaling ansatz of Eq.~\eqref{eq:Ansatz}.
The search for the best scaling exponents $\alpha(r)$ and $\beta(r)$ is done separately under the assumption that they are independent.
This is coherent with our scaling ansatz. 
In practice, one obtains $\alpha(r)$ from the asymptotic slope of the distributions (i.e., the initial dependence on $k$ before the peloton) and $\beta(r)$ from a power law fit $n^{\beta(r)}$ to $k_b(n; r)$ versus $n$.
Our initial findings, based only on two values of $r$, reveal that the exponents have only a mild dependence on $r$ and in particular that $\beta$ seems to be close to $0.3$.  In view of our small datasets, it is not expected that the numerical values of $(\alpha,\beta)= (1.87,0.285)$ used in Fig.~\ref{fig:peloton} are the absolute best scaling exponents.
A complete analytical justification of our scaling ansatz and a derivation of the expected values of the exponents are still lacking.
However, the mere existence of a scaling behavior provides useful estimates of how the degrees of the leaders scale with network size.
This is a crucial information when one is interested in the statistics of the extremes, both in theory \cite{Krapivsky2002} and application \cite{Albert2000}.
This calls for a more extensive study beyond the scope of the present contribution.

%==============================================================================
\section{Parameter estimation}
\label{appendix:parameters}
%==============================================================================
This Appendix presents our parameter estimation method.
The problem is simply stated: We are given an empirical network of $N$ nodes, and an assignment of its nodes in $g$ overlapping communities.
A number of statistics are associated to the network--communities pair: the node membership distribution, the community size distribution, and the internal degree distribution of each communities.
Our task is to identify the parameters $(p,q,r,N,s)$ which will generate synthetic network--communities pairs whose statistics are as close as possible to the statistics of the empirical dataset.
Because the final network size $N$ and basic community size $s$ are both automatically determined by the empirical dataset, this amounts to identifying the optimal value of three free parameters: $p \in [0,1]$  (size of communities),  $q \in [0,1]$  (memberships of nodes), and $r \in [0,\infty)$  (density of communities).

One could be tempted to fit these three free parameters simultaneously, especially since the algorithm of Sec.~\ref{subsec:complete_model-coupling} integrates the local growth model (parametrized by $r$) and SPA (parametrized by $p,q$).
Three observations indicate that this is not necessary.
First, it is clear that $q$ only determines the number of communities to which an average node belongs, a quantity that has no bearing on the internal connectivity of a community.
We can therefore fit this parameter independently of $r$.
Second, the introduction of $\epsilon$ [see Eq.~\eqref{eq:epsilon}] allows us to treat $p$ and $r$ independently, even though both parameters are related to the rate of growth of communities.
That is, we can always obtain a distribution of community sizes of exponent $\gamma_S(p,s)$ and simultaneously generate communities of average asymptotic degree $\langle k \rangle = 2(1+r)$.
Only the value of $\epsilon$---a nonphysical parameter---changes from one set of parameters to the other.
Third, the coupling between $p$ and $q$ is already understood: Changes in  the  value of $p$ mostly affect the size distribution, and changes in  the  value of $q$ mostly affect the membership distribution \cite{HebertDufresne2011,HebertDufresne2012}.
We use the word ``mostly'' because these parameters are independent if $s=1$ (always the case in this study), but there exists a weak coupling if $s>1$; the interplay between the two parameters is then prescribed by 
Eqs.~\eqref{eq:gammaN} and \eqref{eq:gammaS}.
In what follows, we will explain how to fit $p,q$ and $r$ independently from one another, starting with $p$ and $q$.
Note that all \textit{estimated} values of the parameters will be affixed with a caret: $\hat{p}$, $\hat{q}$, and $\hat{r}$.

%~~~~~~~~~~~~~~~~~~~~~~~~~~~~~~~~~~~~~~~~~~~~~~~~~~~~~~~~~~~~~~~~~~~~~~~~~~~~~~
\subsection{Community structure estimators}
\label{appendix:parameters-community_structure}
%~~~~~~~~~~~~~~~~~~~~~~~~~~~~~~~~~~~~~~~~~~~~~~~~~~~~~~~~~~~~~~~~~~~~~~~~~~~~~~
The estimates $\hat{p}$ and $\hat{q}$ are obtained directly from the memberships and size distributions of the empirical data.
We first assume that these distributions are pure power llaws and use a systematic method to extract their exponents by likelihood maximization \cite{Clauset2009}.
We then find a first set of values for $\hat{p}_0$ and $\hat{q}_0$ by inverting Eqs.~\eqref{eq:gammaN} and \eqref{eq:gammaS}.
Because neither the empirical nor the modeled distribution are pure power laws, these values act as first approximations; 
small perturbations ($\Delta \leq 0.1$) to $\hat{p}_0$ and $\hat{q}_0$ can increase the quality of the fit.
We select the estimates $\hat{p}$ and $\hat{q}$ that minimize the difference between the CCDF of the empirical and the simulated distributions.

%~~~~~~~~~~~~~~~~~~~~~~~~~~~~~~~~~~~~~~~~~~~~~~~~~~~~~~~~~~~~~~~~~~~~~~~~~~~~~~
\subsection{Density estimators}
\label{appendix:parameters-density}
%~~~~~~~~~~~~~~~~~~~~~~~~~~~~~~~~~~~~~~~~~~~~~~~~~~~~~~~~~~~~~~~~~~~~~~~~~~~~~~
There exist many methods to fit $r$ to the empirical data.
We will focus on two simple ones.
The first is a straightforward least square estimator (LSE); it compares the distance between the observed average degree $\langle k (n) \rangle$ in groups of size $n$ and the analytical prediction of Eq.~\eqref{eq:ER_link_creation_rate} for $\langle k (n;r) \rangle = 2L(n;r)/n$.
By minimizing the distance over all $r$, one obtains the estimate $\hat{r}$.
The other method is a simple likelihood maximization (MLE) which  relies on the results of the rate equations of Sec.~\ref{subsec:internal_model-description_mean_field}, i.e., on the internal degree distribution, parametrized by the community size $n$.
Let $\{ k_{i,\ell}\}$ be the sequence of internal degrees of a real network, where $i$ refers to node $i$ and $\ell$ is the index of a community of node $i$.
Assuming uncorrelated communities, the log-likelihood $\mathcal{L}$ that $r$ was used to generate the sequence $\{ k_{i,\ell}\}$ is then
\begin{equation}
    \mathcal{L}(r|\{ k_{i,\ell} \}) = \sum_{\ell} \sum_{i\in\ell} \log[p({k_{i,\ell}}|n_\ell, r)]\; ,
    \label{eq:log_likelihood_r}
\end{equation}
where $p({k_{i,\ell}}|n_\ell, r)$ is the probability of finding a node of degree $k_{i,\ell}$ in a community of size $n_\ell$, if the growth ratio equals $r$. 
This probability is obtained by integrating Eq.~\eqref{eq:master_equations_internal_by_size}, with the initial condition  $\vec{p}(t_0)=(p_0=0, p_1=2/3, p_2=1/3)$.
We select the estimate $\hat{r}$ that maximizes Eq.~\eqref{eq:log_likelihood_r}.

%~~~~~~~~~~~~~~~~~~~~~~~~~~~~~~~~~~~~~~~~~~~~~~~~~~~~~~~~~~~~~~~~~~~~~~~~~~~~~~
\subsection{Bias of the density estimators}
\label{appendix:parameters-correction}
\begin{figure}
    \centering
    \includegraphics{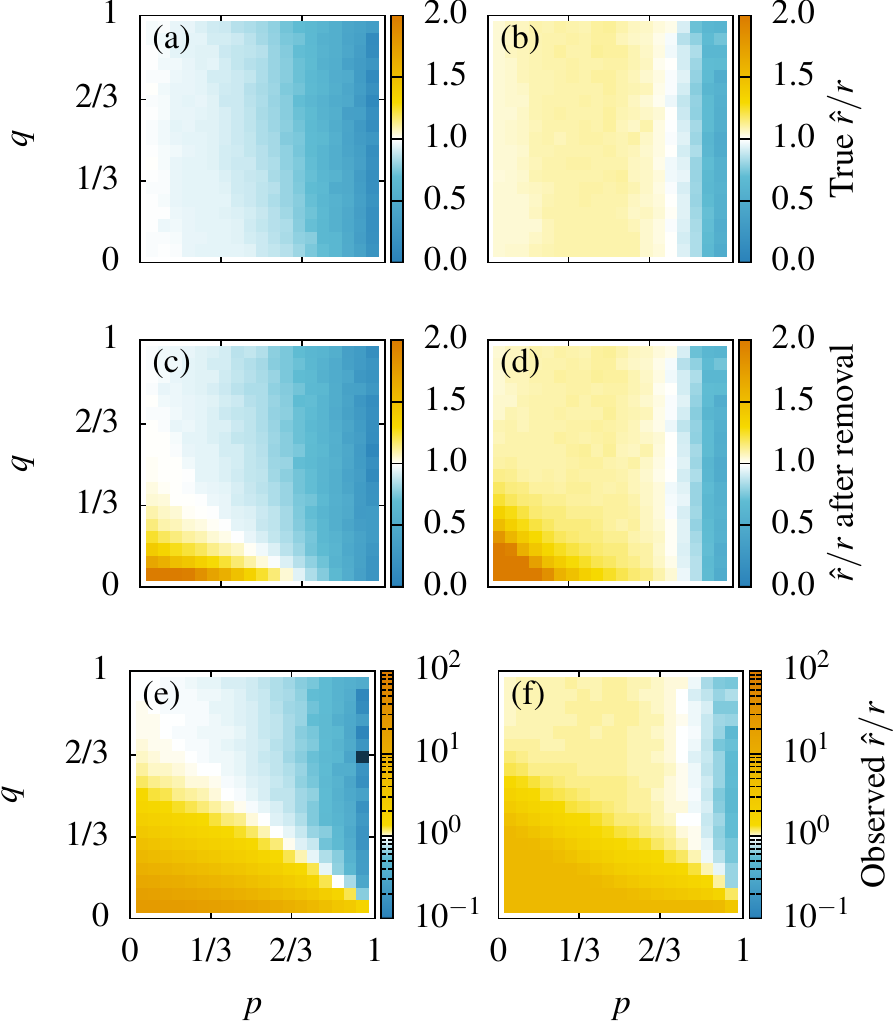} 
    \caption{(color online)
        Accuracy of the least squares [(a), (c), and (e)] and maximum log-likelihood [(b), (d), and (f)] estimators of $r$, in small networks generated by SPA+ with $s=1$, $N=10\,000$ and $r=2$.
        Each sub-figures display the ratios of the estimates $\hat{r}$ to the value of $r$ used to create the data, for all pairs of $(p,q)$, in different test cases (see text of Sec.~\ref{appendix:parameters-correction_bias} for details).
        The cases are as follows:
        [(a) and (b)] estimators computed using the \emph{pure} internal  degree distributions,
        [(c) and (d)] estimators computed using the \emph{collapsed} internal  degree distributions,
        [(e) and (f)] estimators computed using the \emph{collapsed and overlapping} internal  degree distributions.
        A perfect match is color coded in white, whereas under- (over-)  estimates appear in shades of blue (yellow).
        Numerical experiments yield qualitatively similar figures for different values of $N$ and $r$.
        However, the magnitude of the bias is a function of $N$ and of $r$ (see Tables \ref{table:scaling_of_bias_a} and \ref{table:scaling_of_bias_b}).
    }
    \label{fig:finite_size}
\end{figure}

%~~~~~~~~~~~~~~~~~~~~~~~~~~~~~~~~~~~~~~~~~~~~~~~~~~~~~~~~~~~~~~~~~~~~~~~~~~~~~~
There exists three sources of bias for $\hat{r}$: the distribution of community sizes, the redundant memberships discussed in Sec.~\ref{subsec:complete_model-finite}, and the presence of overlap in empirical networks.
In this section, we delineate these effects and introduce a simple correction mask that circumvents the bias.
We use the following procedure to quantify this bias: We construct a number of SPA+ networks and obtain clean matrices of internal degrees $\{ k_{i,\ell}\}$.
By ``clean'', we mean that we do not collapse redundant memberships into single memberships (see Sec.~\ref{subsec:complete_model-finite}), and we do not take overlap into account.
Then, we gradually introduce effects which are present in real systems, and establish how each effect influences the estimate $\hat{r}$.
The numerical results of this investigation are displayed in Fig.~\ref{fig:finite_size} and Tables \ref{table:scaling_of_bias_a}--\ref{table:scaling_of_bias_b}.

% - - - - - - - - - - - - - - - - - - - - - - - - - - - - - - - - - - - - - - -
% Bias of the density estimator: Figure
% - - - - - - - - - - - - - - - - - - - - - - - - - - - - - - - - - - - - - - -
% - - - - - - - - - - - - - - - - - - - - - - - - - - - - - - - - - - - - - - -
% Bias of the density estimator: Table
\begin{table}
    \caption{Mean relative bias $\hat{r}/r$ in networks of $N=5\,000$ nodes.}
    \centering
    \begin{tabular}{l|ccc|ccc}\hline\hline
        Case                              & \multicolumn{3}{c|}{LSE} & \multicolumn{3}{c}{MLE} \\ \hline
        $r$                               & $2$    & $4$  & $8$  & $2$  & $4$  & $8$  \\ \hline
        \textit{Pure}                     & 0.75& 0.78& 0.84& 1.01& 0.99& 1.01\\
        \textit{Collapsed}                & 0.80& 0.81& 0.86& 1.09& 1.05& 1.05\\
        \textit{Collapsed and overlapping}\textsuperscript{a}        & 2.02& 1.50& 1.23& 1.11& 1.03& 1.05\\\hline\hline
    \end{tabular}
    \label{table:scaling_of_bias_a}
    \caption{Mean relative bias $\hat{r}/r$ in networks of $N=10\,000$ nodes.}
    \begin{tabular}{l|ccc|ccc}\hline\hline
        Case                              & \multicolumn{3}{c|}{LSE} & \multicolumn{3}{c}{MLE} \\ \hline
        $r$                               & $2$    & $4$  & $8$  & $2$  & $4$  & $8$  \\ \hline
        \textit{Pure}                     & 0.77& 0.80& 0.85& 1.01& 0.99& 1.00\\
        \textit{Collapsed}                & 0.82& 0.83& 0.88& 1.09& 1.06& 1.04\\
        \textit{Collapsed and overlapping}
        \footnote{We excluded some points from the average $\langle ­\hat{r}/r \rangle$, because the estimates $\hat{r}$ lied outside of the search ranges $\hat{r}\in[0,50]$ (LSE) $\hat{r}\in[0,12]$ (MLE) for $r=4$ and $r=8$. The excluded points are (LSE) those who satisfy $p+q < 0.15$ [a small lower right triangle in the $(p,q)$ space] and  (MLE) those who satisfy $p<0.2$ or $q<0.2$ [left or bottom edge in the $(p,q)$ space].}
        & 2.30& 1.65& 1.32& 1.12& 1.05& 1.04\\\hline\hline
    \end{tabular}
\label{table:scaling_of_bias_b}
\end{table}
% - - - - - - - - - - - - - - - - - - - - - - - - - - - - - - - - - - - - - - -

%~~~~~~~~~~~~~~~~~~~~~~~~~~~~~~~~~~~~~~~~~~~~~~~~~~~~~~~~~~~~~~~~~~~~~~~~~~~~~~
\subsubsection{Effect of community size}
\label{appendix:parameters-correction_size}
%~~~~~~~~~~~~~~~~~~~~~~~~~~~~~~~~~~~~~~~~~~~~~~~~~~~~~~~~~~~~~~~~~~~~~~~~~~~~~~
The estimators are first calibrated on \emph{pure internal structures} [Figs~\ref{fig:finite_size}(a) and \ref{fig:finite_size}(b)].
In this regime, we do not transform the matrices of internal degrees.
It corresponds to the case where communities are directly generated by the model of Sec.~\ref{sec:internal_model}.
The quality of the estimate $\hat{r}$ depends on $p$, through its effect on the distribution of community sizes.
As $p$ increases, the inference task becomes harder, because communities are smaller and mostly live in the fully connected regime, where there are few discriminating features (large ranges of $r$ yield similar internal degree distributions).
The LSE performs best when $p\approx 0$: SPA+ generates only a few extremely large communities, and the internal degree within these communities falls neatly on the plateau of $\langle k (n) \rangle$.
The MLE performs relatively well across a wide range of values of $p$, but we nonetheless observe a positive bias when $0.05 <p < 0.75$: For these values of $p$, SPA+ generates many communities in the intermediate size range, where the mean-field description of the local model is known to be numerically inaccurate (see Fig.~\ref{fig:internal_degree_distribution}).
We also note that there is a noticeable variation of the ratio $\hat{r}/r$ for fixed values of $p$ in the case of the LSE.
This variation is due to changes in the maximum community size: If $q$ is large, then the network quickly reaches the target number of nodes, and the largest communities fall in the linear regime of $\langle k (n) \rangle$.

%~~~~~~~~~~~~~~~~~~~~~~~~~~~~~~~~~~~~~~~~~~~~~~~~~~~~~~~~~~~~~~~~~~~~~~~~~~~~~~
\subsubsection{Effect of redundant memberships}
\label{appendix:parameters-correction_finite}
%~~~~~~~~~~~~~~~~~~~~~~~~~~~~~~~~~~~~~~~~~~~~~~~~~~~~~~~~~~~~~~~~~~~~~~~~~~~~~~
The next case of interest is that of the \emph{collapsed internal structures} [Figs~\ref{fig:finite_size}(c) and \ref{fig:finite_size}(d)].
It is obtained by merging redundant memberships into single entities, and then removing the resulting self-loops and parallel links (see Sec.~\ref{subsec:complete_model-finite}).
As a result of this procedure, communities that contain redundant copies of a same node decrease both in size and number of links.
This leads to denser communities on average.
These effects are only significant at very low values of $p$ and $q$, i.e., for parameters that yield highly redundant communities.
Redundant memberships have been shown to account for a vanishingly small fraction of all memberships when the number of communities is large \cite{HebertDufresne2012}.
However, our numerical experiments show that, for the LSE, the effects of this source of bias do not decrease with system size for extreme values of $(p,q)$---in fact, they increase slightly (see Tables~\ref{table:scaling_of_bias_a} and \ref{table:scaling_of_bias_b}).
This is because the effect of redundant memberships is more prominent in large communities (the most valuable communities for estimating $\hat{r}$), which are more frequent when the network is larger.

%~~~~~~~~~~~~~~~~~~~~~~~~~~~~~~~~~~~~~~~~~~~~~~~~~~~~~~~~~~~~~~~~~~~~~~~~~~~~~~
\subsubsection{Effect of overlap}
\label{appendix:parameters-correctionoverlap}
%~~~~~~~~~~~~~~~~~~~~~~~~~~~~~~~~~~~~~~~~~~~~~~~~~~~~~~~~~~~~~~~~~~~~~~~~~~~~~~

A significant bias is introduced when one does not assign links to specific communities.
This is what we call the \emph{collapsed and overlapping structures}, where links increase the density of \emph{all} the communities to which they belong, rather than a single one.
This  final case encompasses all the biases, and makes use of the information that should be recovered by means of a perfect community detection algorithm.
As shown by our results, the bias is more pronounced in the significantly overlapping regime $p<q$,  where communities grow \emph{slower} than the node reservoir.
Again, our numerical experiment show that the effects of this source of bias increase slightly with community size (see Tables~\ref{table:scaling_of_bias_a} and \ref{table:scaling_of_bias_b}).

%~~~~~~~~~~~~~~~~~~~~~~~~~~~~~~~~~~~~~~~~~~~~~~~~~~~~~~~~~~~~~~~~~~~~~~~~~~~~~~
\subsubsection{Bias removal mask}
\label{appendix:parameters-correction_bias}
%~~~~~~~~~~~~~~~~~~~~~~~~~~~~~~~~~~~~~~~~~~~~~~~~~~~~~~~~~~~~~~~~~~~~~~~~~~~~~~
Since most overlapping community detection algorithms do not explicitly assign links, we are often placed in the ``collapsed and overlapping'' case.
We use the following modeling procedure to account for the bias: (i) obtain the parameters $(\hat{p},\hat{q})$ that best model the community structure, (ii) compute an initial estimate $\hat{r}_0$ of the strength of the internal connectivity of communities, and (iii) finally obtain a corrected estimate $\hat{r}$, as $\hat{r}=\hat{r}_0/M(\hat{p},\hat{q};N)$.
The correction $M(\hat{p},\hat{q};N)$ is the value of the bias removal mask for networks of $N$ nodes at point $(\hat{p},\hat{q})$.
Since the mask depends on the network size $N$ (see Tables~\ref{table:scaling_of_bias_a} and \ref{table:scaling_of_bias_b}), it is computed for each network separately.
In practice, we obtain $M(\hat{p},\hat{q};N)$ by first generating a number of SPA+ networks of $N$ nodes with fixed parameters $(\hat{p},\hat{q}, r=r_M)$.
We have found that the final results are almost independent on the precise value of $r_M$; we have used $r_M=2$ but $r_M=\hat{r}_0$ is an equally good choice.
We then extract $\hat{r}_M$ from the collapsed and overlapping communities (averaged over the number of SPA+ networks realizations) and take $M(\hat{p},\hat{q};N) = \hat{r}_M/r_M$.
This bias removal mask allows us to generate networks with mean internal degrees on a $\langle k (n) \rangle$ curve that resemble the empirical data.
We use the MLE because it is more stable with respect to changes in $r_M$.

%\bibliographystyle{apsrev4-1}
%\bibliographystyle{abbrv}
% \bibliography{dunbar_initialsonly}
% \end{document}

%

\end{document}